\documentstyle[11pt,epsfig]{article}

\def\ee{\end{equation}}
\def\be{\begin{equation}}
\def\eea{\end{eqnarray}}
\def\bea{\begin{eqnarray}}
\def\ea{\end{eqnarray}}
\def\ba{\begin{eqnarray}}\def\eeas{\end{eqnarray*}}
\def\beas{\begin{eqnarray*}}

\begin{document}
\title{The Goldstone bosons in the pairing Hamiltonian:  
the path integral approach} 
\author{M.B. Barbaro$^1$, A. Molinari$^1$, F. Palumbo$^2$ 
and M.R. Quaglia$^3$\\ 
${}^1$ Dipartimento di Fisica Teorica --- Universit\`a di Torino 
\\ 
Istituto Nazionale di Fisica Nucleare --- Sezione di Torino\\ 
 Torino ---Italy \\ 
${}^2$ INFN --- Laboratori Nazionali di Frascati \\ 
P.O. Box 13, I-00044 Frascati --- Italy\\ 
${}^3$ Dipartimento di Fisica --- Universit\`a di Genova\\ 
Istituto Nazionale di Fisica Nucleare --- Sezione di Genova\\ 
Genova --- Italy\\} 
\date{} 
\maketitle

\begin{abstract} 
As a first step to derive the IBM from a microscopic 
nuclear hamiltonian, 
we bosonize the pairing hamiltonian in the framework of the path integral 
formalism respecting both the particle number conservation and the Pauli 
principle. Special attention is payed to the role of the Goldstone bosons.  
We construct the saddle point expansion which reproduces the 
sector of the spectrum associated to the  
addition or removal of nucleon pairs. 
\end{abstract} 
 
\section{Introduction} 
 
The problem of relating the interacting {\em boson} model (IBM),  
so successful in interpreting the low-energy nuclear phenomenology,  
to an underlying {\em fermionic} nuclear hamiltonian  
has been attacked in a number of papers (see in particular  
\cite{Gin,Cas,Iac}). 
 
In principle, a possible procedure to deal with it could be 
\begin{enumerate} 
\item 
to derive an effective interaction in a chosen model space, 
\item 
to express it in terms of pairing, 
quadrupole and other forces, 
\item  
to bosonize the model hamiltonian thus obtained. 
\end{enumerate} 
 
Less ambitiously one could start by assuming from the outset a multipole  
expansion of the effective interaction in the chosen model space.  
 
Following these steps one should be able to relate the parameters 
appearing in the IBM to those 
occurring in the fundamental nucleon-nucleon interaction and the bosonic 
operators to the original fermionic ones. 
 
A vast literature exists concerning the problem of mapping the 
fermionic model space into the bosonic one~\cite{Klein}. However  
it appears to us that the actual realization of  
these mappings still deserves futher analyses, especially in connection to
the nature of the interacting bosons, 
namely whether they are Goldstone bosons or not. 
Let us comment on why the Goldstone bosons should play a role  
in nuclear structure and how they relate to some of the bosons of the IBM.   
In this connection, we remind that  
underlying the IBM is the recognition that the nuclear interaction is  
attractive enough to form pairs of nucleons. In low energy phenomena these   
conserve their identity inside nuclei, thus becoming the relevant degrees of  
freedom for the collective nuclear levels. 
The analogy between these pairs of nucleons and the   
Cooper pairs of superconductivity is clearly suggestive and indeed 
the simplified version of the BCS theory of superconductivity represented
by the so-called pairing hamiltonian was applied to atomic nuclei long ago
and was dealt with the Bogoliubov-Valatin (BV) transformation.

The pairing hamiltonian, in its simplest form, governs the dynamics of pairs
of nucleons moving in a mean field and coupled to zero angular momentum
(referred to as $s$-bosons in the IBM).
The great success of the IBM stemmed from the introduction of pairs of 
nucleons coupled to angular momentum two as well, the $d$-bosons, which can
also be treated through a generalized BV transformation. 
In all cases BV does not conserve the number of particles.
Although elaborated methods have lately been devised to preserve the
number of particles~\cite{Klein}, 
it is of importance to realize that this feature
of the BV transformation is connected to the central concept lying at the
core of superconductivity, namely the spontaneous breaking of the global gauge
invariance related to the particle number conservation.
Thus in a superconducting system there must exist 
an associated Goldstone boson.
Now in the IBM nuclei are indeed viewed as superconducting systems, 
even if only approximately, since they are finite systems. 
As a consequence, the IBM   
should embody at least one Goldstone boson. Other Goldstone bosons should of  
course appear if other symmetries, like the rotational one,  
are (approximately) spontaneously broken.    
 
In infinite systems the Goldstone fields carry 
specific signatures: indeed they live in the coset space of the  
spontaneously broken group $U(1)$ of the global gauge transformation
$e^{iq\Lambda/\hbar}$, $q$ being the electric charge,
with respect to the unbroken subgroup $Z_2$ consisting of the gauge 
transformations associated with $\Lambda=0$ and $\Lambda=\pi \hbar/q$
and display only derivative interactions. 
If these distinctive features survive in finite systems, the identification of
 Goldstone bosons becomes of relevance not only for a  
deeper understanding of 
the bosonization mechanism, but also for a convenient choice of
the variables.
One could indeed choose bosonic fields not living  
in the coset space, but at the price of rendering the formalism quite  
cumbersome, as exemplified by the chiral physics \cite{Wei}. 
 
Motivated by the above heuristic considerations, in this paper 
we present a new investigation of the program outlined at the beginning  
of this Introduction,  
limiting ourselves to confront the third point, namely to consider
only the pairing interaction. 
Yet, since, in a model space, the pairing interaction is an important component
of any realistic effective interaction~\cite{Dean},  
then our work may also be viewed as  
a first step in the derivation of the IBM. 
 
We will use the path integral formalism where the problem of  
superconductivity is readily solved independently from which quantum  
numbers the Cooper pairs carry \cite{Wei}. 
The basic question we address is to which extent the features of  
superconductivity in an infinite system, in particular the signatures of  
the Goldstone fields, survive in a finite system. Our main result is  
that indeed these signatures are exactly preserved in the pairing model.
Accordingly we argue that the extension of the latter to include higher 
multipolarities of the force should be easily feasible,  
as confirmed by recent work \cite{Pal4}. 
 
The plan of the present paper is as follows.  
 
In Section \ref{pairham} we discuss from the point of view illustrated above
the well known spectrum of the pairing hamiltonian, which 
is characterized by the two quantum numbers $n$ (the number of pairs) and $s$ 
(the number of broken pairs), naturally related to two types of excitations.  
Those associated with $n$ relate to 
the addition or removal of a correlated pair of nucleons and can be viewed 
as Goldstone bosons stemming from the (almost) broken  
conservation of particle number. Those related to $s$, namely to 
the breaking of pairs (seniority excitations), we view as corresponding to 
the Higgs particles.

In Section \ref{lattice} we derive the euclidean path integral formulation and 
introduce auxiliary bosonic fields via the Hubbard-Stratonovitch 
transformation. 
 
In Sections \ref{sec:saddle} and \ref{first} we set up a saddle point  
expansion of the effective action of the auxiliary fields. 
The expansion parameter turns out to be the inverse of the energy 
\be 
M= \frac{g\Omega}{2} \ , 
\label{M} 
\ee 
$g$ being the strength of the pairing force and $\Omega=j+1/2$ the pair 
degeneracy of a single-particle level with angular momentum $j$ 
(for the sake of simplicity we shall consider the case of one level only).  
Within such an expansion we succeed in reproducing the excitation energies  
associated with the addition or removal of pairs 
of nucleons. We do not explore however the seniority excitations, which 
are not easily accomodated in the framework of our expansion since 
their energies are of the order of $M$.  
A reasonable estimate, based on the available nuclear phenomenology~\cite{BM},
yields $M\geq$3 MeV in the region of the Sn isotopes. 
As anticipated, our analysis shows that the pairing model encodes  
in a striking way the  
basic features of spontaneous symmetry breaking in an infinite system:  
{\em indeed 
the field which describes the Goldstone excitations lives in the coset 
of the U(1) symmetry related to particle number conservation with respect to 
the conserved $Z_2$ subgroup and displays only derivative couplings.} 
 
Then we investigate two paths for selecting a sector 
of a given number of nucleons: one based on the chemical potential  
(Section \ref{chemicalpotential}) and 
the other on a projection operator (Section \ref{projection}).  
Since at zero temperature, 
even in the presence of the chemical 
potential, the number of particles does not fluctuate, 
the two formalisms lead exactly to the same result. 
 
In Section \ref{bosons} we derive the  hamiltonian of the $s$-bosons. In
the present case, since the spectrum of the pairing model is known,
this hamiltonian has been determined in a simpler, direct way by
Talmi~\cite{Talm1}. This work, however, ignores the Pauli principle, 
which, therefore, must be added {\it a posteriori} as an ad hoc prescription.
Moreover our approach has the merit that it
can be applied as well when forces of higher multipolarity
are active, thus opening the way to the microscopical derivation of
the IBM. In Section \ref{concl} we present our conclusions and outlooks. 
 
For the sake of completeness, in concluding this Introduction, 
we mention two approaches to the problem of bosonization attempted in the past. 
 
The first, also limited to the pairing interaction, was based on the  
use of even Grassmann variables in the generating functional
\cite{Pal1,Pal2}, but substantial difficulties  
were met in its extension to include the quadrupole interaction. 
 
The second was carried out by Mukherjee and Nambu~\cite{Nam}  
who explored in depth the connection between 
the BCS theory of superconductivity and the IBM. These authors, linearizing 
in the frame of the mean field approach a nuclear BCS hamiltonian embodying 
a contact four-nucleon interaction, accounted for second order corrections.
They were thus able to derive a bosonic Hamiltonian expressed as a sum 
of Casimir operators, hence qualitatively of the IBM type. 
However, these authors actually explored an infinite homogeneous 
system and dealt with finite nuclei only in some approximate schemes,  
thus failing to enforce the particle number conservation.  
Moreover their approach appears hardly suitable for a realistic  
derivation of the IBM model, in particular as far as the fermion-boson mapping 
is concerned. 
 
\section{The excitations of the pairing hamiltonian as Goldstone and 
Higgs bosons} 
\label{pairham} 
 
According to the framework outlined in the Introduction we 
consider the schematic pairing Hamiltonian, which accounts for  
much of the physics of the low energy spectra of nuclei.  
Our aim is to establish a connection between its well-known spectrum 
and the (almost) spontaneous breaking of the particle number conservation,  
which entails the existence of Goldstone and Higgs modes. 
 
In its simplest version the pairing hamiltonian  
describes a system of interacting 
bounded identical nucleons living in one single particle level 
of angular momentum $j$ and reads 
\be 
\hat H= \sum_{m=-j}^j e_m \hat \lambda^{\dag}_m \hat \lambda_m -g \hat A^{\dag} \hat A 
\label{HP} 
\ee 
where 
\be 
\hat A = \sum_{m>0}(-1)^{j-m} \hat \lambda_{-m} \hat \lambda_m\ , 
\ee 
${\hat \lambda}^{\dag}$, ${\hat \lambda}$ are the usual  
creation and annihilation nucleon operators,  
$m$ is the third component of the angular momentum $j$,  
the $e_m$ are the (negative) single-particle energies and $g$ is the strength  
of the pairing force. 
For the sake of simplicity we set $e_m=e$ independent of $m$. 
In the Conclusions we will mention how and when  
the level dispersion can be accounted for \cite{Bar02}. 
 
In this paper we shall consider an even number of identical nucleons only. 
In such a case the energy spectrum is given by the well-known  
formula~\cite{Row} 
\be 
E_{n,s} 
=2  e n-gn(\Omega-n+1)+gs(\Omega-s+1)\ ,\,\,\,n \ge s 
\label{ePb} 
\ee 
where $n$ is the number of pairs and 
$s$ the pair seniority 
\footnote{According to our definition the pair seniority quantum number $s$  
is half the usual seniority $v$ and corresponds to the number of pairs not 
coupled to angular momentum $J=0$, and, as such, blind to the action of the 
pairing force.}. Clearly (\ref{ePb}) holds valid for  
\be 
n \le \Omega 
\label{nlimit} 
\ee 
and not only for $s\leq n$, but for 
\be
s\leq \Omega-n
\ee
as well, as a consequence of the Pauli principle.  

In an infinite system the energy of the Goldstone bosons vanishes with
the associated quantum number. This does not occur in a finite system, but, to
the extent that the energy spectrum of the latter displays a pattern 
similar to that of an infinite system, it should exhibit two quite different 
energy scales.
Actually the excitation energies associated with both the quantum 
numbers $n$ and $s$ appear to be of order $g\Omega$.  
However the Goldstone nature of the energy spectrum associated with
the quantum number $n$ is clearly apparent when one considers the excitations 
with respect to the minimum. This, if $n$ is viewed as a 
continuous variable, 
occurs for 
\be 
\nu_0= { 1 \over 2}(\Omega + 1) - { e \over g}    
\,. 
\label{n0} 
\ee 
In fact, since $n$ assumes only integer values, the minimum of (\ref{ePb}) 
takes place for  
\be 
n_0=[\nu_0], 
\ee 
$[...]$ meaning integral part.  
Introducing then the shifted quantum number 
\be 
\nu=n-n_0 
\label{nu} 
\ee 
 (\ref{ePb}) becomes 
\be 
E_{n_0+\nu,s} = g\nu^2 + 2g \nu (n_0 - \nu_0 ) - g n_0 ( 2\nu_0 - n_0 )    
+g s(\Omega-s+1)\,. 
\label{exact} 
\ee 
Now we see  that the addition (or removal) of one pair of nucleons with respect to  
the ground  
state requires an energy of order $g$: this is the energy of the Goldstone  
boson. Instead the energy required to break a pair, the seniority energy,  
is of order $g\,\Omega$: this is the energy of a Higgs boson. 
 
We should now point out that using the physical values for $\Omega$, 
$g$ and $e$ appropriate, for example, to the $Sn$ nucleus (these can be taken 
from \cite{BM}, \cite{Bes}), one would obtain a value for  
$\nu_0$ corresponding to an unphysical nucleus. 
However the excitation energies of both modes,  
measured with respect to the minimum (see eq.~(\ref{exact})) 
are {\it essentially independent} of the single-particle energy $e$.  
Hence our argument, although heuristic, remains qualitatively correct. 
The real justification of its validity will be given in the next Section. 
 
Finally, since in the pairing hamiltonian the degeneracy $\Omega$ is fixed by 
the model space, the excitation energies related to the quantum numbers  
$\nu$ and $s$ can be predicted once two conditions are  
chosen in order to fix the parameters $e$ and $g$.

\section{The generating functional} 
\label{lattice} 
 
As is well-known \cite{Negele} the path integral must be 
evaluated in its discretized form. 
The discretized euclidean action of our system is 
\be 
S=\tau \sum_{t=-N_0/2}^{N_0/2-1} \left\{ -g \overline A(t) A(t-1)+ 
\sum_{m=-j}^j \left[\overline \lambda_m (t) \left( \nabla_t^+ 
+e\right) \lambda_m(t-1)\right]  
\right\}\ , 
\ee 
where $\tau$ is the time spacing, $N_0$ the number of points on the  
time lattice, 
\be 
\left( \nabla_t^{\pm} f\right)(t)=  
\pm\frac{1}{\tau}\left[ f(t\pm 1)-f(t)\right] 
\ee 
and  
\be 
Z=\int \left[d\overline \lambda d\lambda\right] e^{-S} 
\ee 
the generating functional. 
Moreover  
$\lambda$, $\bar\lambda$, 
$A=\sum_{m>0}(-1)^{j-m}\lambda_{-m}\lambda_m$ 
and $\bar A=\sum_{m>0}(-1)^{j-m}\bar\lambda_{m}\bar\lambda_{-m}$ 
are Grassmann variables. 
We remind that the fermion fields must satisfy antiperiodic boundary conditions 
in time. 
 
Now, to cast the action in a form convenient for the saddle point expansion,  
we perform a number of manipulations: their role will be illustrated when 
appropriate. 
First we shift the time label in the variables $\lambda$ (but not $\overline\lambda$) 
according to 
\be 
\lambda(t-1) \rightarrow \lambda (t) 
\label{shift} 
\ee 
in order to have $\overline A$ and $A$ with the same time argument. 
This yields for the action the expression 
\be 
S^I=\tau \sum_{t=-N_0/2}^{N_0/2-1}  
\left\{ -g \overline A(t) A(t)+\sum_{m=-j}^j \left[ 
\overline \lambda_m (t) \left( \nabla_t^+ 
+e\right) \lambda_m(t) \right]  
\right\}\,. 
\ee 
Carrying out next the Hubbard-Stratonovitch transformation, we get the new action 
 \bea 
S^{II}&=& \tau \sum_{t=-N_0/2}^{N_0/2-1} \Bigg\{  
g \overline \eta(t) \eta(t) +  
g\overline \eta (t) A(t) +g \eta(t) \overline A(t) 
\nonumber\\ 
&+& 
\sum_{m=-j}^j 
\left[ \overline \lambda_m (t)  
\left( \nabla_t^+ +e\right) \lambda_m(t) \right] 
\Bigg\}\,. 
\label{SII} 
\eea 
Clearly the auxiliary fields $\overline\eta$ and $\eta$ should satisfy periodic 
boundary conditions. 
 
Finally we introduce the Goldstone field $\theta$ through the 
polar representation~\cite{Wei} for the auxiliary fields  
\begin{eqnarray} 
\eta &=& \sqrt{\rho} e^{2i\theta},  \,\,\,\,\, \overline\eta = \sqrt{\rho} 
 e^{-2i\theta}\,. 
\label{polar} 
\end{eqnarray} 
The field $\rho$ has been placed under square root to avoid the jacobian
which would otherwise appear. 
Notice that for this change of variable to be  one to one 
(with the only exception of the point $\rho=0$), $\theta$ must vary in the 
range $0 \le \theta < \pi$. Hence the field $\theta$ 
lives in the coset space of the (almost) broken symmetry group $U(1)$  
of particle conservation with respect  to the unbroken subgroup $Z_2$, 
as appropriate to a Goldstone field~\cite{Wei}. 
From the periodic boundary conditions 
for the $\eta$-field periodic boundary conditions for $\rho$ and $\theta$ 
follow as well.
 
We note now that, after the transformation (\ref{polar}), the 
$\theta$ field appears in the action (\ref{SII}) with 
non-derivative couplings whereas the Goldstone field should display
only derivative couplings. 
However the former can be eliminated 
introducing the following transformation on the nucleon fields:  
\be 
\lambda_m = e^{i\theta} \psi_m,   
\,\,\,\,\, \overline\lambda_m = e^{-i\theta}  
\overline\psi_m\,. 
\label{transf} 
\ee 
 
As a consequence of the above transformation the following operators 
\bea 
q^{\pm} 
 &=& \exp ( \mp i \theta) \nabla_t^{\pm} \exp ( \pm i \theta) \pm e ~,
\eea 
whose matrix elements read
\bea 
\left(q^+\right)_{t_1 t_2} &=& \frac{1}{\tau}\left[ 
\exp\left\{i\tau\left(\nabla_t^+\theta\right)_{t_1}\right\}\delta_{t_2,t_1+1} 
-\delta_{t_1 t_2}\right]+e \delta_{t_1 t_2} 
\\ 
\left(q^-\right)_{t_1 t_2} &=& \frac{1}{\tau}\left[\delta_{t_1 t_2}- 
\exp\left\{i\tau\left(\nabla_t^+\theta\right)_{t_2}\right\}\delta_{t_2,t_1-1} 
\right]-e \delta_{t_1 t_2}\,, 
\eea 
will appear in the action.
The $\theta$ field appears only in these operators and therefore  
under derivative, as appropriate to a Goldstone field. 
 
Hence (\ref{SII}) can be recast in the form 
\begin{eqnarray} 
S^{III}&=& \tau \sum_{t=-N_0/2}^{N_0/2-1}  
\Bigg\{ g \rho + \sum_{m>0} \left[ 
\overline \psi_m q^+ \psi_m +\overline \psi_{-m} q^+ \psi_{-m}  
\right. 
\\ 
&& 
\left. 
+g \sqrt{\rho} (-1)^{j-m}  
\left(\overline \psi_m \overline \psi_{-m} 
+\psi_{-m}\psi_m\right) 
\right]\Bigg\} 
\nonumber\\ 
&& 
=\tau \sum_{t=-N_0/2}^{N_0/2-1}  
\Bigg\{ g \rho + \sum_{m>0} \left[\left(\overline\psi_m+ 
g\sqrt{\rho}(-1)^{j-m} \psi_{-m}(q^+)^{-1}\right)q^+ 
\right. 
\nonumber 
\\ 
&&\times\left.\left(\psi_m+g\sqrt{\rho}(-1)^{j-m}  
(q^+)^{-1}\overline\psi_{-m}\right) 
-g^2\rho\psi_{-m}(q^+)^{-1}\overline\psi_{-m}+  
\overline\psi_{-m}q^+\psi_{-m} \right]\Bigg\}\,. 
\nonumber 
\end{eqnarray} 
 
Now we first integrate over the fermionic fields 
$\overline\psi_m$ and $\psi_m$ for a given positive $m$: 
this yields $\mbox{Det}(q^+)$, independent of $m$. 
Likewise, performing the integration over  
$\overline\psi_m$ and $\psi_m$ for a given negative $m$, we get 
$\mbox{Det}(q^++g^2\rho[(q^+)^{-1}]^T)$, $T$ meaning the transpose operation, 
again $m$-independent. 
Lumping the two results together and exploiting the relation 
\be 
(q^+)^T=-q^- 
\label{qT} 
\ee 
we find for the fermionic functional integration the result 
\be 
\left[\mbox{Det}\left(-q^-q^++g^2\rho\right)\right]^\Omega\,. 
\ee 
Thus, disregarding here and in the following all the field independent factors, 
we get for the generating functional the following expression 
 \be 
Z=\int_0^\infty \left[d\rho \right] \int_0^{\pi} \left[ d\theta \right] \exp (-S_{eff})\,, 
\label{Zeff} 
\ee 
with 
\be 
S_{eff} = \tau \sum_t g \rho  - \mbox{Tr} \ln \left(-q^-q^+ +g^2\rho\right)\ . 
\ee 
Note that the argument of the logarithm is symmetric, but not hermitian. 
The trace must be taken over the quantum number $m>0$ and the time. 
The $U(1)$ symmetry is now non linearly realized 
in the invariance of $S_{eff}$ under the substitution 
\be 
\theta \rightarrow \theta+\alpha\,, 
\label{invariance} 
\ee 
with $\alpha$ time independent.  
 
\section{The saddle point} 
\label{sec:saddle} 
 
In this Section we look for a minimum of $S_{eff}$ at constant fields:  
hence only the time-independent component of the $\rho$ field, to be referred  
to as $\overline \rho$, will enter into the effective action. 
We start by defining 
\bea 
M &= & \sqrt{e^2+g ^2 \overline \rho} 
\label{eqM} 
\eea 
and 
\bea 
P^{-1} &= & -\nabla_t^+ \nabla_t^- + e \left(\nabla_t^+ - \nabla_t^-\right) + 
M^2 \nonumber\\ 
& =& -(1-e \tau)\  \Box + M^2\,, 
\label{P-1} 
\eea 
where $\Box=\nabla_t^+\nabla_t^-$. 
Notice that $e\tau$ cannot be neglected with respect to one.  
Indeed in our calculations we will first 
perform the limit $N_0 \rightarrow \infty$ and then we shall let 
$\tau \rightarrow 0$. 
The effective action at constant fields reads then 
\bea 
\overline S_{eff}&=&  
\tau \sum_t g \overline \rho - \mbox{Tr} \ln P^{-1}\,. 
\eea 
The trace is conveniently evaluated in the Fourier representation, yielding 
\bea 
\overline S_{eff} 
&=& \tau N_0 g \overline \rho 
-\Omega \sum_{n_0=-N_0/2}^{N_0/2-1} \ln \left[  
4(1-e \tau) \sin^2 {\pi\over N_0}(n_0+1/2) +\tau^2 M^2  
\right]\,, 
\nonumber\\ 
\label{29} 
\eea 
where the antiperiodic boundary conditions of the nucleon 
fields have been taken into account. Converting the sum into an integral we get 
\bea 
\overline S_{eff} 
&=&N_0 \tau \left\{g \overline \rho -\frac{2\Omega}{\tau}\ln\left[\frac{1}{2} 
\left(\tau M+\sqrt{4(1-e \tau)+\tau^2 M^2}\right)\right] 
\right\} 
\nonumber\\ 
&=& 
N_0 \tau\left(g\overline\rho+\Omega e-\Omega M\right) 
+ {\cal O}(\tau^2)\,. 
\eea 
Notice that the piece $\Omega e$ stems from the term $\tau e$ in 
(\ref{29}). 
The minimum of $\overline S_{eff}$ occurs for 
\be 
M=\frac{g\Omega}{2}\,, 
\ee 
which is independent of $e$ and $\overline\rho$. 
Inserting the above into (\ref{eqM}) one gets 
for the value $\overline\rho_0$ of  
$\overline\rho$ at the minimum the expression 
\be 
\overline\rho_0=\frac{\Omega^2}{4}-\frac{e^2}{g^2}\,, 
\label{ovrho0} 
\ee 
so that $\overline S_{eff}$ at the minimum is 
\be 
S_0 
=N_0\tau\left(-\frac{M\Omega}{2}-\frac{e^2}{g} 
+\Omega e\right)\,. 
\label{S0} 
\ee 
Although the values for $\Omega$, $g$ and $e$ appropriate for the 
$Sn$ isotopes would lead to a negative $\overline\rho_0$,  
actually in selecting a given nucleus (see Section \ref{chemicalpotential}) 
the replacement $e\to \epsilon=e-\mu$ ($\mu$ being the chemical potential) 
should be performed. 
When this is done, as it will be shown in Section \ref{chemicalpotential}, 
$\overline\rho_0$ turns out indeed to be positive, as it should be, in the 
physical range $1\le n\le \Omega$, attaining its maximum value for 
$n\simeq \Omega/2$, namely for a half-filled level.

\section{The saddle point expansion} 
\label{first} 
  
To perform this expansion we start by defining the fluctuation of the  
 $\rho$-field according to 
\be 
\rho = \overline{\rho}_0 + r 
=\overline{\rho}_0\left(1 + \frac{r}{\overline{\rho}_0}\right) 
\label{eq:34} 
\ee 
and by noticing that the generating functional (\ref{Zeff}) now reads 
\be 
Z=\int_{-\overline\rho_0}^\infty \left[d r \right]  
\int_0^{\pi} \left[ d\theta \right] \exp (-S_{eff})\,. 
\label{Zeff1} 
\ee 
Now two cases should be considered:
\begin{itemize}
\item[a)] $\overline\rho_0$ sufficiently large: then the functional  
integral defining $Z$ becomes gaussian and an expansion in 
$r/\overline\rho_0$ can clearly be performed.
Actually, as it will be later shown (see formula (\ref{nOn}) below), 
$\overline\rho_0$ is indeed large when 
$n\sim\Omega/2$, namely when the level where the pairs live is far from being 
fully occupied or empty;
\item[b)] $\overline\rho_0$ small, which occurs for $n\simeq 1$ or
$n\simeq\Omega$ (see again formula (\ref{nOn}) below). 
In this case the shift
in (\ref{eq:34}) is absent and the $\rho$ field acts only through its 
fluctuations, which are small, thus assuring the validity of the expansion.
\end{itemize}
 
To  proceed further we rewrite $S_{eff}$ in the form 
\be 
S_{eff} = \tau \sum_t g \left(\overline{\rho}_0 + r \right) + \mbox{Tr} \ln P 
-  \mbox{Tr} \ln \left[ 1\!\!1 + P \left (R_1 +R_2\right) \right]\ , 
\label{sefff}
\ee 
where 
\be 
R_1 = -q^-q^+ + ( \nabla_t^+ + e) ( \nabla_t^- - e) 
\ee 
and 
\be 
R_2 = g^2 r 
\,. 
\ee 
We set then  
\be 
S_{eff}=\sum_{r=0}^\infty S_r\,, 
\label{40}
\ee 
the term $S_0$ being the saddle point contribution, given by 
(\ref{S0}). This grows like $ \Omega^2$, however it contains  
also a term of order $\Omega$ and a term of order one, which should be kept
if an expansion in powers of $1/\Omega$ is sought for.
It seems to us, however, more convenient to stick to the definition 
(\ref{P-1}) for the operator $P$ and to compute the further contributions
to the expansion (\ref{40}) (the quantum fluctuations) by developing
the logarithm (\ref{sefff}): the terms thus obtained are naturally organized
in powers of $M^{-1}$.
It is worth reminding that this expansion does not break 
the $U(1)$ invariance. 
 
In the following we shall confine ourselves to evaluate, 
in addition to the first order terms,
those quadratic in $\nabla_t\theta$ and $r$.  
 
\subsection{First order contributions} 
\label{firstorder}
 
These contributions stem from the term linear in $r$ and 
from the first term in the expansion of the logarithm, hence 
\be 
S_1 = \tau g \sum_t r_t  
-  \mbox{Tr} \left[ P \left (R_1 +R_2\right ) \right]\,. 
\label{S1} 
\ee 
The explicit computation of the second term on the rhs 
of the above yields 
\begin{eqnarray} 
-\mbox{Tr} \left( P R_1 \right) 
&=& \frac{\Omega}{\tau^2} \sum_t P_{tt}\Bigg\{ 
\left[ 1 - e^{2i(\theta_{t+1}-\theta_t)}\right] 
\nonumber\\ 
&-& 
\left[ 1 - e^{i(\theta_{t+1}-\theta_t)}\right] 
\left[ \tau^2 M^2-\frac{\tau^2}{P_{tt}}+2(1-e\tau)\right] 
\Bigg\}\,, 
\end{eqnarray} 
where $P_{tt}$ is found to read 
\be 
P_{tt}=\frac{\tau}{2M}\left(1-e\tau+\frac{\tau^2 M^2}{4}\right)^{-1/2}\,. 
\ee 
By expanding the exponentials up to second order in $\theta$ we get 
\begin{eqnarray} 
-\mbox{Tr} \left( P R_1 \right) = \frac{\Omega}{2M} 
\left[1+\left(M+\frac{3}{2}e\right)\tau\right] \tau 
\sum_t \frac{\left(\theta_{t+1}-\theta_t\right)^2}{\tau^2} + {\cal O}(\tau^3) \,. 
\end{eqnarray} 
 
The contribution arising from the third term on the 
rhs of (\ref{S1}) turns out to be 
\bea 
-\mbox{Tr}(PR_2) &=& - \Omega g^2 \sum_t P_{tt} r_t 
\,. 
\label{TrPR2} 
\eea 
Notably this contribution, linear in $r$, is canceled by the first term  
in $S_1$, owing to the equation for the minimum of the action  
$\overline S_{eff}$.  
The cancellation holds to the order ${\cal O}(\tau^2)$, which is the  
approximation we keep in our analysis and in obtaining
the equation for the action minimum. 
 
In conclusion, for the first order contribution to the action we get 
\be 
S_1=-\mbox{Tr} (PR_1) = {1 \over g} \left[ 1 + (M+\frac{3}{2} e)\tau \right]   
\tau \sum_{t=-\infty}^{\infty} \theta (-\Box ) \theta\,, 
\ee 
where $\sum_t$ runs from $-\infty$ to $\infty$ since we let $N_0\to\infty$ in  
evaluating $P$. 
We note that these contributions are of order $\Omega$ and 1, namely are  
$O(1/\Omega)$ with respect to those of the saddle point.

\subsection{Second order contributions} 
 
We have seen in the previous subsection that all the terms linear in $r$ cancel out:  
hence the $r$-integration remains undefined.  
Our aim now is to ascertain  whether  
the surviving terms in $r$ stabilize the action.  
 
Among these we consider the contributions 
arising from the second term in the expansion of the logarithm. They read 
\be 
S_2=\frac{1}{2}  
\mbox{Tr} \left (P R_1\right )^2 
+\mbox{Tr} \left( P R_1 P R_2\right ) 
+\frac{1}{2} \mbox{Tr} \left (P R_2\right )^2. 
\ee 
In the above the first term is $r$-independent, the second is linear in $r$ 
and the third one is quadratic. Therefore, for the present purpose,  
it is sufficient to evaluate the latter. For this we have found  
\be 
\frac{1}{2} \mbox{Tr}(PR_2)^2 = \frac{g^4\Omega}{2}  
\sum_{t t_1} r_t \left[P_{t t_1}\right]^2 r_{t_1}\,. 
\ee 
Hence the integral over $r$ is well defined.  
 
In conclusion we remark that,   
as it will be seen in the following Sections, in order to obtain
the Goldstone bosons energies we must find out how they depend upon the
single particle energy $e$. 
For this purpose we have to perform in the integral expressing the  
generating functional $Z_1$ (associated with the action $S_1$) 
the $\theta$-integration, which appears to be 
gaussian, but actually it is not, because $\theta$ is compact. Yet we can  
choose $\nabla_t \theta$ as a new integration variable, thus rendering 
the integral gaussian. 
We then get 
\be 
-\frac{1}{N_0\tau}\ln Z_1= 
\frac{3}{4} e + \frac{M}{2}= 
\frac{3 e+g\Omega}{4}\,. 
\ee 
 
We notice that this contribution stems from the term
$(M + 3e/2)\tau$ in $S_1$, which is irrelevant because it vanishes in the
formal continuum limit.

\section{Fixing the particle number by the chemical potential} 
\label{chemicalpotential} 
 
In this Section we apply the saddle point expansion to a specific nucleus  
using the method of the chemical potential.  
For this purpose we replace $e$ with   
\be 
\epsilon=e-\mu\ , 
\ee 
$\mu$ being the chemical potential. Its value is fixed according to 
\be 
<\hat N>= \frac{1}{N_0\tau} \frac{\partial}{\partial \mu} \ln 
Z=  -\frac{1}{N_0 \tau} \frac{\partial}{\partial \epsilon} \ln Z\,, 
\label{numero} 
\ee 
where  
\be 
\hat N=\sum_{m=-j}^j\hat\lambda_m^\dagger\hat\lambda_m 
\ee 
is the particle number operator. Since, however,  
we shall let $N_0\to\infty$ with $\tau$ constant, which corresponds to the
limit of vanishing temperature $T=1/(N_0\tau)$,
we are allowed to replace $<\hat N>$ with $2n$. 
We also notice that because $M$ does not depend upon $e$ (see eq. (\ref{M})),  
it does not depend on $\mu$ either.  
So Eq.~(\ref{numero}) becomes 
\be 
n = { 1 \over 2N_0 \tau} { \partial \over \partial \epsilon}  
(S_0-\ln Z_1)= 
-\frac{\epsilon}{g}+\frac{\Omega+3/4}{2}\,, 
\ee 
which gives  
\be 
\mu= g \left( n - \Omega/2-3/8\right) + e 
\label{mu1} 
\ee 
for the chemical potential. 
Hence, in the presence of the chemical potential, the energy of the system 
becomes 
\bea 
E_{n,0} &=& \frac{1}{N_0 \tau}(S_0-\ln Z_1) + 2 \, \mu n  
\nonumber\\ 
&=&  2 e n - g n (\Omega - n+3/4)+\frac{g}{8}\left(5\Omega+\frac{9}{8}\right) 
\,. 
\label{E1mu} 
\eea 
We thus see from the above that in our approach the excitation spectrum of  
the pairing hamiltonian is reproduced with good accuracy.  
On the other hand the ground state energy differs from the exact 
value $-g[(\Omega+1)/2-e/g]^2$  
by the quantity $(3 g \Omega + g -e)/4$, which corresponds to 
a relative error of order $1/\Omega$. 
 
We conclude this Section by further examining the issue, already addressed
in the beginning of Section \ref{first}, of the validity of 
our expansion.
For this purpose it is of importance to assess the size of $\overline\rho_0$. 
To this aim we replace in (\ref{ovrho0}) $e$ by $\epsilon$ and use (\ref{mu1}),
dropping the term -3/8 in the round brackets on the RHS, 
thus getting 
\be 
\overline\rho_0=n\left(\Omega-n\right)\,. 
\label{nOn} 
\ee 
Now, when $n\sim\Omega/2$, then the single particle energy $e$ almost 
coincides with the chemical potential $\mu$. In such a situation $\epsilon$ 
is almost vanishing and, from (\ref{nOn}), $\overline\rho_0\simeq\Omega^2/4$. 
This large value corresponds to the situation when
the level where the pairs live is neither fully filled 
nor almost empty. 
On the other hand $\overline\rho_0$ attains its lowest value when $n=1$
or $n=\Omega$. It is remarkable that even in these cases, where an expansion
in $1/\overline\rho_0$ cannot be performed, our approach still yields
the correct excitation spectrum of the system.
 
\section{Fixing the particle number by the projection operator} 
\label{projection} 
 
Owing to the importance of properly fixing the particle number $n$,  
in this Section we address the problem through an alternative procedure, 
namely by introducing in the path integral the projection operator 
\be 
{\cal P}_n=\int_{-\pi}^{+\pi} \frac{d\alpha}{2\pi} e^{-i (\hat N-2n)\alpha}\ . 
\label{Pn} 
\ee 
Using then the variables (\ref{transf}) and performing the  
Hubbard-Stratonovitch transformation as previously done, we get for the  
generating functional the expression 
\be 
Z^{(n)}=\int_{-\pi}^{+\pi}  \frac{d\alpha}{2\pi} 
\int\left[d\lambda d\overline\lambda d\psi d\overline\psi d\eta d\overline \eta \right] 
e^{-S^{(n)}} 
\label{Z} 
\ee 
where 
\bea 
&&{S}^{(n)} 
=\tau \sum_{t=-N_0/2}^{N_0/2-1}  
\Bigg\{ g \rho + \sum_{m>0} \left[ 
\overline \psi_m {q}^+_\sigma \psi_m +\overline \psi_{-m} {q}^+_\sigma  
\psi_{-m} \right. 
\nonumber\\ 
&& 
\left. 
+g \sqrt{\rho} \left(\overline \psi_m \overline \psi_{-m} 
+\psi_{-m}\psi_m\right) 
\right]\Bigg\}-2 N_0\tau(\sigma-e) n 
\nonumber\\ 
\eea 
(remember that the label $n$ indicates the number of pairs), having defined 
\be 
\sigma= e+\frac{i\alpha}{N_0\tau} 
\ee 
and 
\bea 
q^\pm_\sigma 
 &=& \exp ( \mp i \theta) \nabla_t^\pm \exp ( \pm i \theta) \pm \sigma \,.
\eea 
  
Next we carry out the integration over the fermionic degrees of freedom, 
getting for the partition function the expression
\be 
Z^{(n)}\propto 
\int_{e-\frac{i\pi}{N_0\tau}}^{e+\frac{i\pi}{N_0\tau}} d\sigma  
\int  \left[d\rho d\theta\right] 
\exp [-S^{(n)}_{eff}(\sigma,\rho,\theta)] 
 \ , 
\label{Zz} 
\ee 
being  
\bea 
S^{(n)}_{eff} &=&  
-2N_0\tau(\sigma-e)n +\tau \sum_t g \rho   
- \mbox{Tr} \ln \left[-{q}^-_\sigma {q}^+_\sigma  +g^2 \rho \right] 
\,. 
\label{Seffective} 
\eea 
 
At constant fields  
(\ref{Seffective}) simplifies to 
\be 
{\overline S^{(n)}}(\sigma,\overline\rho) 
=N_0\tau\left[g\overline\rho+2n e+(\Omega-2n) \sigma 
-\Omega\sqrt{\sigma^2+g^2\overline\rho}\,\right]\,, 
\label{barSn} 
\ee 
which is stationary when 
\be 
\frac{\partial {\overline S^{(n)}}(\sigma,\overline\rho)} 
{\partial \overline \rho}=  
N_0\tau g \left( 1-\frac{\Omega g}{2\sqrt{\sigma^2+ g^2 \overline \rho}} 
\right)=0  
\ee 
and 
\be 
\frac{\partial {\overline S^{(n)}}(\sigma,\overline\rho)} 
{\partial \sigma}  
=N_0\tau \left(-\frac{\Omega\sigma}{M} + \Omega -2n\right)=0\,. 
\ee 
The solutions of the above equations read 
\bea 
\overline\rho_0&=& n(\Omega-n) 
\eea 
and 
\bea 
\sigma_0&=&\frac{g}{2}(\Omega-2n) 
\label{sigmazero} 
\,. 
\eea 
Finally the effective action (\ref{barSn})  
at the minimum $\sigma_0 $ turns out to be 
\be 
\overline S^{(n)}(\sigma_0,\overline\rho_0)=  
N_0\tau E_0^{(n)} = N_0\tau [2ne -gn (\Omega-n)]\ , 
\label{S0projector} 
\ee 
which differs from the zero seniority spectrum of the pairing hamiltonian.
Indeed the latter has $\Omega+1$, rather than $\Omega$, inside the round 
bracket on the RHS of (\ref{S0projector}).
 
Note that the action (\ref{barSn}) is an analytic  
function of $\sigma$ inside an integration path deformed to encompass 
the saddle point $\sigma_0$.  
This path goes  
from $e-i \pi/N_0\tau$ to $\sigma_0-i \pi/N_0\tau$ along a straight  
line parallel to the real axis, then from $\sigma_0-i \pi/N_0\tau $ 
to $\sigma_0+i \pi/N_0\tau$ along a straight line parallel to the imaginary  
axis and finally it goes back from $\sigma_0+i \pi/N_0\tau$ to  
$e +i \pi/N_0\tau$. When $N_0\to \infty$, the contributions coming from the  
paths parallel to the real axis cancel each other, while the one parallel  
to the imaginary axis vanishes: hence there are no corrections to the saddle  
point contribution. 
 
Next, with the aim of checking the results obtained in the framework of
the chemical potential method, we evaluate, using the projection operator, 
the first  order correction in the saddle point expansion 
in $\rho$ and $\theta$. 
For this scope we set, as in (\ref{eq:34}),  
\be 
\rho = \overline \rho_0 + r\,. 
\ee 
The action (\ref{Seffective}) can then be recast as follows 
\bea 
S_{eff}^{(n)}\simeq 
\overline S_0^{(n)}(\sigma_0,\overline\rho_0) + S_1^{(n)} 
\eea 
where 
\be 
S_1^{(n)} = N_0 \tau gr  
-\mbox{Tr} [P(\sigma_0,\overline \rho_0) (R_1^{(n)}+R_2^{(n)})] 
\label{s1n} 
\ee 
with 
\be 
R_1^{(n)}=  
-[q^-_\sigma  + (\sigma -\sigma_0)][q^+_\sigma -(\sigma -\sigma_0)]+  
(\nabla_t ^+ +\sigma_0) (\nabla_t ^- -\sigma_0) 
\ee 
and 
\be 
R_2^{(n)}= g^2 r\,. 
\ee   
 
The contribution  
\be 
\mbox{Tr} [P(\sigma_0,\overline \rho_0) R_2^{(n)}] 
= g^2  
\mbox{Tr} P(\sigma_0,\overline \rho_0)r 
= g^2 \frac{\tau}{g}\sum_t r  
\ee 
cancels the first term in (\ref{s1n}). Hence the latter simply becomes 
\be 
S_1^{(n)}= - 
\mbox{Tr} [ P(\sigma_0,\overline \rho_0) R_1^{(n)}]\,. 
\ee 
A calculation, similar to the one carried out in Section \ref{firstorder}, 
yields then  
\bea 
-\mbox{Tr} \left[ P R_1^{(n)} \right] &=&  
\frac{\Omega}{2M} 
\left[1+\left(M+\frac{3}{2}\sigma_0\right)\tau\right] \tau 
\sum_t \frac{\left(\theta_{t+1}-\theta_t\right)^2}{\tau^2}\,. 
\eea 
Using now the expression (\ref{sigmazero}) for $\sigma_0$ we get 
\be 
S_1^{(n)}=  
{1 \over g} \left[ 1 + g\left(\frac{5}{4}\Omega-\frac{3}{2}n\right)\tau \right] 
 \tau \sum_t \theta (-\Box ) \theta\,
\ee 
and by performing the $\theta$-integration (again using (\ref{sigmazero}))
we finally obtain the first order energy  
\be 
E_0^{(n)}+E_1^{(n)}=2ne -gn\left(\Omega-n+\frac{3}{4}\right)+\frac{5}{8}g  
\Omega\,,  
\ee 
which coincides with the
excitation spectrum obtained with the chemical potential. 
 
\section{The hamiltonian of the s-bosons}
\label{bosons}

To complete our program (restricted, we remind, to the pairing 
potential) we derive below the bosonic hamiltonian corresponding to 
our effective action.
The most general, particle conserving, quartic hamiltonian 
for a system of $s$-bosons, confined lo live in one single particle level,
reads in normal form 
\be
H(\hat{b}^{\dagger}, \hat{b}) =h \hat{b}^{\dagger} \hat{b}  + 
v \hat{b}^{\dagger}  \hat{b}^{\dagger}  \hat{b}\hat{b} ,
\label{Hbos}
\ee
$\hat{b}^{\dagger},\hat{b}$ being bosonic 
creation-annihilation operators acting in 
a Fock space. They satisfy canonical commutation relations. 
The values of the parameters $h,v$ were obtained in ref.~\cite{Talm1} 
in such a way to yield the pairing Hamiltonian spectrum. 
Of course the Hamiltonian thus obtained, being intrinsically bosonic, 
patently violates the Pauli principle and therefore the condition $n< \Omega$
should be added {\it a posteriori}, when using (\ref{Hbos}) in describing a 
system of fermions. Here, to show how this condition naturally emerges instead 
in our framework, we obtain the parameters $h$ and $v$ with our methods.  
To this purpose we write the path integral associated to (\ref{Hbos}), namely
\be
Z=\int [db^*db] \exp (-S) ,
\ee
where  
\be
S= \tau \sum_{t=1}^{N_0} 
\left\{ b^*_{t+1} \nabla_t b_t + H(b_{t+1}^*,b_t) \right\}
\ee
and the $b^*,b$ are holomorphic variables satisfying periodic boundary 
conditions in time. 
We now compare the above to our effective action.

To this end we introduce the polar representation 
\be
b = \sqrt{\rho} \exp (i \theta),\,\,\,b^* = \sqrt{\rho} \exp (- i \theta)
\ee
in terms of which the generating functional 
and the action read
\ba
Z&=&\int_0^{\infty}[d \rho] \int_{- \pi}^{\pi}[d \theta] \exp (-S)
\\
S &=& \tau \sum_t \Big\{ \sqrt{\rho_{t+1}} \exp( -i \theta_{t+1}) \nabla_t \, 
\left[ \sqrt{\rho_t} \exp( i \theta_t)
\right] 
\nonumber\\
&+&  H\left[ \sqrt{\rho_{t+1}} \exp( - i \theta_{t+1}), 
\sqrt{\rho_t} \exp( i \theta_t\right] \Big\}.
\eea
Now we again 
look for a minimum of $S$ at constant fields and perform a saddle point
expansion. The calculation is basically the same as the one previously
developed and hence will not be reported here. 
We only remind that such an expansion holds valid only
for $n < \Omega$, as it follows from the positivity of $\bar\rho_0$, 
given in eq.~(\ref{nOn}): thus in our approach such a condition, far from 
being artificial, is necessarily implied by the formalism itself.
We only quote the result of the comparison with our effective action: 
it yields
\be
h=2e-g\Omega- g/4  ,\,\,\,v=g .
\ee
Obviously the considerations following (\ref{40}) hold valid as well here.
Thus $h$ will be affected by an error of order $1/M$.
Indeed for the Hamiltonian (\ref{Hbos}) to reproduce exactly the pairing spectrum 
it must be 
\be
h=2e-g\Omega ,\,\,\,v=g .
\ee

Finally, it is of importance to stress once more that the above discussion 
can be generalized to include other types of bosons, as, for example,
those appearing in the quartic hamiltonian of the IBM model,
respecting basic symmetries like particle number conservation,
rotational invariance, etc.
It is then clear that our approach opens the way to microscopically deduce
the Arima-Iachello model:
in this case to fix the coupling constants one should again
write the corresponding path integral and compare the resulting bosonic 
action with the effective one found extending the procedure developed in 
this paper to a 
fermionic hamiltonian including forces of higher multipolarity. 

\section{Conclusions} 
\label{concl} 
 
In this paper we have carried out an investigation concerning the  
possibility of a systematic bosonization of a realistic 
nuclear hamiltonian for the description of the low-lying sector of the 
nuclear spectrum.  
Our study is admittedly preliminary since it is limited to the pairing 
interaction, which is of course only a component (although an important 
one) of an effective interaction pretending to be realistic. 
However, having overcome the main difficulty we expected, namely  
the one of going from an infinite to a finite system,  
we are now confident to be able to solve the bosonization problem in the 
presence of other types of interaction~\cite{Pal4}. 
 
Our approach is based on the concept
of symmetry breaking and on the related properties of the 
interaction among the bosonic fields: indeed
this framework is the most suitable for deriving, rather than assuming,
a model like the IBM one.
 
To develop our scheme we have used the path integral formalism  
because of the large flexibility it allows both in  
choosing and in dealing with the variables appropriate to the problem. 
This has lead us to deduce an asymptotic expansion for the system's spectrum
in the parameter $M^{-1}=2/g\Omega$.  
 
The euclidean path integral clearly encompasses the whole Fock space  
of the system. 
To deal with a specific nucleus a given number of pairs must be selected 
(a procedure not to be confused, of course,
with the projection of the particle number when this is violated).  
In our approach a definite particle sector has been chosen using
both the chemical potential and the projection operator methods: 
the two turn out to be completely equivalent.  
 
Notably our expansion to first order reproduces with good accuracy  
the energy of the pair addition and removal modes (or, in the language of the 
IBM, of the $s$-bosons). 
Moreover the requirement that $\overline\rho_0$ 
should be positive entails the inequality $n<\Omega$
(see again eq.~(\ref{nOn})), thus implementing
the action of the Pauli principle. 
This is conceptually important for the consistency of our scheme, 
which respects both the Pauli principle and the  
particle number conservation.  
Worth noticing is that this crucial feature is  
absent in the framework developed in \cite{Talm1,Talm2},  
but correctly dealt with in many treatments based on the mapping  
procedure~\cite{Gin,Klein}. 
  
Our expansion cannot account for the seniority excitations, 
whose energies, in the framework of the pairing Hamiltonian (\ref{HP}), 
are larger than $M$: hence they should be separately treated. 
For this scope clearly an analysis of the excitations associated with the
field $\rho$ should be performed.
In this connection it is worth reminding 
that in the scheme of the BV transformation
these modes are described in terms of quasi-particles whose energies,
as well-known, are expressed in terms of the gap $\Delta$.
In an infinite system $\Delta$ is associated with the order parameter by
setting a non-zero vacuum expectation value for the pair (Cooper) field
and it signals the onset of the superconducting phase; 
in a finite system it measures the strength of the mean field (referred to
as the ``pair field'') which, in the BV scheme, 
linearizes the pair interaction.

Finally we should observe that in our analysis the impact on the 
excitation spectrum stemming from the removal of the  
degeneracy of the levels where the pairs are sitting and of the higher  
order terms in the saddle point expansion has not been explored.  
Concerning the first issue, if the spacing between the single particle  
energy levels is small with respect to $M$, it can easily be accounted for
within the present perturbative scheme.  
Indeed this approach has already been pursued in ref.~\cite{Bar02}. 
Actually the Sn isotopes are well suited for such perturbative treatment 
since here the distance between the  single particle energies of concern 
appears to be much smaller than $M$.   
 
Concerning the second point, we can only say that it would certainly be both
interesting and important to examine in more depth our asymptotic expansion.

\section*{Acknowledgements} 
We wish to acknowledge useful conversations with Prof. N. Lo Iudice and
G. Pollarolo.

\end{document}